\documentstyle[aps,multirow,amssymb]{revtex}

\begin{document}

\title{Swelling-Collapse Transition of Self-Attracting Walks}

\author{A.~Ordemann$^{1,2}$, G.~Berkolaiko$^{1,3}$, S.~Havlin$^{1}$,
        and~A.~Bunde$^{^2}$}

\address{$^1$Minerva~Center~and~Department~of~Physics,
         Bar-Ilan~University,
         52900~Ramat-Gan, Israel}
        
\address{$^2$Institut~f\"ur~Theoretische~Physik~III,
         Justus-Liebig-Universit\"at~Giessen,
         Heinrich-Buff-Ring~16,   
         35392~Giessen, Germany}

\address{$^3$School~of~Mathematics,
         University~of~Bristol,
         Bristol~BS8~1TW, UK}
         

\maketitle
         
\begin{abstract}  

We study the structural properties of self-attracting walks in $d$ dimensions
using scaling arguments and Monte Carlo simulations. We find evidence for a
transition analogous to the $\Theta$ transition of polymers. Above a critical
attractive interaction $u_{\mathrm{c}}$, the walk collapses and the exponents
$\nu$ and $k$, characterising the scaling with time $t$ of the mean square
end-to-end distance $\langle R^2\rangle \sim t^{2 \nu}$ and the average number
of visited sites $\langle S\rangle \sim t^k$, are universal and given by
$\nu=1/(d+1)$ and $k=d/(d+1)$. Below $u_{\mathrm{c}}$, the walk swells and the
exponents are as with no interaction, i.e. $\nu=1/2$ for all $d$, $k=1/2$ for
$d=1$ and $k=1$ for $d \ge 2$. At $u_{\mathrm{c}}$, the exponents are found to
be in a different universality class.

\end{abstract} 


\newpage

In recent years different models of random walks with memory or interaction
have been studied. They can be divided in static~\cite{Domb83} and
dynamic~\cite{Amit83,Sapoz98} models, for an overview we refer to the papers of
Duxbury and Queiroz~\cite{Domb83} and Oettinger~\cite{Amit83}.  Most efforts
concentrated on models with repulsive interactions, in particular self-avoiding
walks (\textit{SAW}), which have been found useful for investigating polymers
in dilute solution. When an attraction term $\exp(-A/T)$, $A<0$, is included, 
the \textit{SAW} model reveals a swelling-collapse transition at the `$\Theta$
point' $T=\Theta$~\cite{deGennes79,Barat95}. In contrast, the likewise
challenging case of random walks with a similar attractive interaction, but
without repulsion, has been less understood. This problem was solved only for
one dimension, while in higher dimensions the results are highly controversial.
Our numerical and analytical study of attractive random walks suggests that
there exists a swelling-collapse transition, too, that is analogous to the
$\Theta$ transition in polymers.

We focus on the dynamic model of self-attracting walks
(\textit{SATW})~\cite{Sapoz98}, where a random walker jumps with probability $p
\sim \exp(nu)$~\cite{negative} to a nearest neighbor site, with $n=1$ for
already visited sites and $n=0$ for not visited sites. The interaction
parameter $u$ is equivalent to $-A/T$ for linear polymers. For $u>0$, the walk
is attracted to its own trajectory~\cite{TSAW}. The structural  behavior of the
walk can be characterized by the mean square end-to-end distance $\langle
R^2(t)\rangle$ and the average number of visited sites $\langle S(t)\rangle$. 
It is expected that these quantities scale with time $t$ as  
\begin{eqnarray}
  \label{generell}
  \hspace*{1cm} \langle R^2(t)\rangle \sim t^{2 \nu} \nonumber \hspace*{6.5cm} (\arabic{equation}\text{a})\\
\text{and} \nonumber \hspace*{15.8cm} \\
   \hspace*{1cm} \langle S(t)\rangle \sim t^{k} \; . \nonumber \hspace*{6.45cm} (\arabic{equation}\text{b})
\end{eqnarray}
\addtocounter{equation}{1}

Earlier analyses for the  \textit{SATW} in two and three dimensions were not
conclusive and the numerical data have been controversially
interpreted~\cite{Sapoz98,Lee98,Reis95,Prasad96}.  While
Sapozhnikov~\cite{Sapoz98} considered the possibility of the existence of a
critical attraction $u_{\mathrm{c}}$ (but his numerical results were not
conclusive), Lee~\cite{Lee98} and Reis~\cite{Reis95} argue strongly against the
existence of $u_{\mathrm{c}}$, since they  find $\nu$ and $k$ continuously
decreasing with $u$.

In this Letter we present scaling arguments and extensive numerical simulations
for $\langle R^2\rangle$ and $\langle S\rangle$ that strongly suggest the
existence of a critical attraction $u_{\mathrm{c}}$ in $d\ge2$, with three
different universality classes for $u > u_{\mathrm{c}}$, $u < u_{\mathrm{c}}$
and $u = u_{\mathrm{c}}$. Below $u_{\mathrm{c}}$, the \textit{SATW} is in the
universality class of random walks, with $\nu = 1/2$  and $k=1$. Above 
$u_{\mathrm{c}}$, the \textit{SATW} collapses and the exponents change to
$\nu=1/(d+1)$ and $k=d/(d+1)$. At the critical point, the exponents are
$\nu_{\mathrm{c}}= 0.40 \pm 0.01$ and $k_{\mathrm{c}}=0.80 \pm 0.01$ in $d=2$
and $\nu_{\mathrm{c}}=0.32 \pm 0.01$ and $k_{\mathrm{c}}=0.91 \pm 0.03$ in
$d=3$~\cite{asymptotic}. The existence of $u_{\mathrm{c}}$ is in striking
similarity to the `$\Theta$ point' phenomenon of linear
polymers~\cite{deGennes79,Barat95} where three different universality classes
for  $T > \Theta$, $T = \Theta$ and $T < \Theta$ exist.
\bigskip   

We used Monte-Carlo simulations to study $\langle R^2(t)\rangle$ and $\langle
S(t)\rangle$. Figure~\ref{3D} shows representative results of $\langle
R^2(t)\rangle$, for several values of $u$ in $d=3$. For large values of $u$,
the curves bend down towards the slope of $2 \, \nu \cong 0.5$, while for small
values of $u$, the curves bend up towards the slope of $2 \, \nu \cong 1$. At
some intermediate critical value $u_{\mathrm{c}} \cong 1.9$, the slope is
approximately $2 \, \nu_{\mathrm{c}} \cong 0.64$. The mean number of visited
sites $\langle S(t)\rangle$ shows similar behavior, with $k \cong 1$ below 
$u_{\mathrm{c}}$, $k_{\mathrm{c}} \cong 0.91$ at $ u_{\mathrm{c}}$, and $k
\cong 0.75$ above $ u_{\mathrm{c}}$. Figure~\ref{nuk}a summarizes the
asymptotic exponents $\nu$ and $k$ as a function of $u$ in $d=3$. We obtained
similar results in $d=2$, the asymptotic values of $\nu$ and $k$ are presented
in Fig.~\ref{nuk}b.  In Table~\ref{table} the values of the exponents are
summarized and compared with the analogous known exponents for the $\Theta$
transition in linear polymers.

In the following we present analytical arguments for the exponents above
criticaltity, which can explain our numerical findings. We assume that for
sufficiently strong attraction $u > u_{\mathrm{c}}$ the grown clusters are
compact, so that the average number of visited sites scales with the rms
displacement $ \langle R(t)\rangle \equiv {\langle R^2(t)\rangle}^{1/2} $ as 
\begin{equation}  
\langle S(t)\rangle \sim {\langle R(t)\rangle}^{d}, \qquad u \gg u_{\mathrm{c}}.
\label{havlin}
\end{equation}
Comparing Eq.~(\ref{generell}) and  Eq.~(\ref{havlin}) yields
\begin{equation}  
k = \nu d, \qquad  \qquad u \gg u_{\mathrm{c}}.
\label{kcompact}
\end{equation}
For sufficiently strong attraction it takes a very long time for the walker to
jump to an unvisited site. Before doing this, the walker diffuses around on the
visited sites, being located with equal probability on any of the cluster
sites. Hence the mean cluster growth rate is proportional to the ratio between
the number of  boundary sites and the total number of the cluster
sites~\cite{Sapoz98,Rammal83}:
\begin{equation}
  \frac{{\rm d} \langle S\rangle}{{\rm d} t}
                \sim \frac{{\langle R\rangle}^{d-1}}{{\langle R\rangle}^{d}} \sim t^{-\nu}.
 \label{sapoz}
\end{equation}
Thus $ \langle S\rangle \sim t^{-\nu +1}$. Combining this 
result  with Eq.~(\ref{generell}b) and~(\ref{kcompact}), we obtain
\begin{eqnarray}
  \label{exponents}
  \hspace*{1cm} \nu = \frac{1}{d + 1} \nonumber \hspace*{7cm} (\arabic{equation}\text{a})\\
\text{and} \nonumber \hspace*{15.8cm} \\
   \hspace*{1cm} k= \frac{d}{d + 1}  \nonumber \hspace*{7cm} (\arabic{equation}\text{b})
\end{eqnarray}
\addtocounter{equation}{1}
for $u \gg u_{\mathrm{c}}$.

Because of universality we assume that these results, which are in agreement
with the exact values $\nu = 1/2$ and $k = 1/2$  in $d=1$~\cite{Prasad96} and
are supported by our extensive  Monte Carlo simulations in $d=2$ and $d=3$, are
valid for all $u > u_{\mathrm{c}}$. Indeed, Fig.~\ref{nuk} suggests that the 
predictions for $u > u_{\mathrm{c}}$  (Eq.~(\ref{exponents})) are approached
asymptotically. We like to note that in $d=2$ the relation $k= \nu d$ also
holds for $u \le u_{\mathrm{c}}$, while in $d=3$ the numerical results yield $k
< \nu d$ for $u < u_{\mathrm{c}}$. Since the mass of the generated clusters
scales like $ M \sim  S \sim { R}^{k/\nu}$, $k/ \nu$ corresponds to the fractal
dimension $d_{\mathrm{f}}$ of the cluster. In $d=2$ the clusters are compact
for all $u$ as $k/\nu = d_{\mathrm{f}}=d$. In $d=3$ they are compact for $u >
u_{\mathrm{c}}$, while for $u < u_{\mathrm{c}}$, the fractal dimension of
clusters generated by simple random walks $d_{\mathrm{f}} =2<d $ is obtained. 
At the criticality, we find $d_{\mathrm{f}} =2.84 \pm  0.25 $, but we cannot
rule out the possibility that $d_{\mathrm{f}} = d $.

To understand the behavior in the critical regime we suggest the following
scaling approach. Guided by Fig.~\ref{3D}, we assume that there exists a
crossover time $t_\xi$ below which the exponent $\nu$ is close to
$\nu_{\mathrm{c}}$ and  above which $\nu$ approaches $1/2$ for $u <
u_{\mathrm{c}}$ and $1/(d+1)$ for $u > u_{\mathrm{c}}$. This suggests the
following scaling relations: 
\begin{eqnarray}
  \label{scaling}
  \hspace*{1cm} R(t) \sim t^{\nu_{\mathrm{c}}} f_{\pm}(t/t_\xi) \nonumber \hspace*{6.3cm} (\arabic{equation}\text{a})\\
\text{and} \nonumber \hspace*{15.8cm} \\
   \hspace*{1cm} S(t) \sim t^{k_{\mathrm{c}}} g_{\pm}(t/t_\xi) \; , \nonumber \hspace*{6.05cm} (\arabic{equation}\text{b}) \\
\text{where} \nonumber \hspace*{15.4cm} \\
   \hspace*{1cm} t_\xi = |u-u_{\mathrm{c}}|^{-\alpha} \; . \nonumber \hspace*{6.4cm} (\arabic{equation}\text{c})
\end{eqnarray}
\addtocounter{equation}{1}
The plus sign refers to $u > u_{\mathrm{c}}$, the minus sign to $u <
u_{\mathrm{c}}$, and the exponent $\alpha$ has to be determined  numerically.
As $t_\xi$ is assumed to be the only relevant time scale, the scaling functions
bridge the short time and the long time regime. To match both regimes, we
require that $f_{\pm}(x) = {\mathrm{const}}$ for $x \ll 1$ ($t \ll t_\xi$),
and  $f_{+}(x) \sim {x}^{1/(d+1) - \nu_{\mathrm{c}}}$,  $f_{-}(x) \sim {x}^{1/2
- \nu_{\mathrm{c}}}$ for $x \gg 1$. Analogous results are expected for
$g_{\pm}(x)$, with $g_{\pm}(x) = {\mathrm{const}}$ for $x \ll 1$, and 
$g_{+}(x) \sim {x}^{d/(d+1) - k_{\mathrm{c}}}$, $g_{-}(x) \sim x^{1 -
k_{\mathrm{c}}}$ for $x \gg 1$. 

To test the scaling theory and to determine the exponent $\alpha$ we 
plotted $\langle R^2(t) \rangle / t_{\xi}^{2 \nu_{\mathrm{c}}}$ and $\langle
S(t) \rangle / t_{\xi}^{k_{\mathrm{c}}}$ as functions of $t/t_\xi$ for several
values of $\alpha$ in $d=2$ and $d=3$. We  obtained the best data collapse for
$\alpha = 5.0 \pm 0.5$ in $d=3$ and  $\alpha = 7 \pm 1$ in $d=2$, which are
shown in Fig.~\ref{collaps3D}a and~\ref{collaps3D}b, respectively. The
excellent data collapse strongly supports the above scaling assumptions. 

\bigskip   

We would like to thank Dmitry Malykhanov for the assistance with the
simulations. Financial support from the German-Israeli Foundation (GIF), the
Minerva Center for Mesoscopics, Fractals, and Neural Networks and the Deutsche
Forschungsgemeinschaft is gratefully acknowledged.


\newpage

\begin{table}[b]
\begin{center}
\begin{tabular}{c c || c | c | c || c | c | c }
 & & \multicolumn{3}{c||}{\textit{RW}} & \multicolumn{3}{c}{\textit{SAW}}  \\ 
 &	& $u < u_{\mathrm{c}}$ & $u = u_{\mathrm{c}}$ & $u > u_{\mathrm{c}}$ & $1/T < 1/\Theta$ & $1/T = 1/\Theta$ & $1/T > 1/\Theta$   \\\hline \hline
\multirow{3}{25mm}{\hspace{10mm} $d=2$}
 & $\nu$ & $1/2$ & $0.40 \pm 0.01$  & $1/3$ & $3/4$ & $4/7$ & $1/2$  \\
 & $k $ & $1 $ & $0.80 \pm 0.01 $ & $ 2/3$ & $ 1 $ & $ 1 $  & $ 1 $ 	  \\ 
 &   	& \multicolumn{3}{c||}{$u_{\mathrm{c}}  =0.88 \pm 0.05$} 	& \multicolumn{3}{c}{$1/\theta_0 =0.65 \pm 0.03$} \\\hline
\multirow{3}{25mm}{\hspace{10mm} $d=3$}
 & $\nu $ & $1/2$ & $0.32 \pm 0.01$ & $1/4$ & $0.59$ & $1/2$ & $1/3$   \\
 & $k $ & $1 $ & $ 0.91 \pm 0.03 $ & $ 3/4$ & $ 1 $ & $ 1 $ & $ 1 $ 		 \\ 
 &  	& \multicolumn{3}{c||}{$u_{\mathrm{c}}  =1.92 \pm 0.03$} 	& \multicolumn{3}{c}{$1/\theta_0 =0.5 \pm 0.03$} \\
\end{tabular}
\end{center}
\caption{Comparison of the exponents $\nu$ and $k$ as well as of the estimated
values for the transition points $u_{\mathrm{c}}$ for random walks
(\textit{RW}) and $\Theta$ for \textit{SAW} on hypercubic lattices. For values
related to the $\Theta$ transition see~\protect\cite{Barat95} and references
therein.}
\label{table} \end{table}

\newpage

\begin{figure}
\caption{ The mean square end-to-end distance $\langle R^2(t) \rangle$
versus $t$ up to $t= 10^8$ timesteps averaged over $1000$ configurations for
each attraction $u=0, \, 1.5, \, 1.9, \, 2.25, \, 4$ in $d=3$. Note that for large values
of $u$ the curves bend down towards the slope of $2 \, \nu= 1/2$,
while for small values of $u$ the curves bend up towards the slope of $2 \, \nu =
1$.}
\label{3D}
\end{figure}

\begin{figure} 
\caption{The values of the exponents $k$ and $\nu$ versus attraction $u$ in (a)
$d=3$  and (b) $d=2$, obtained by a least square fit of the slope of $\ln
\langle R^2(t) \rangle$ and $\ln \langle S(t) \rangle $ versus $\ln t$ for
large t, respectively (see Fig.~\ref{3D}).  Shown are the results for $t=
10^6$~($\bigtriangleup$), $t= 10^7$~($\bigtriangledown$) and $t=
10^8$~($\blacksquare$). Note that for $u > u_{\mathrm{c}}$ and larger $t$ the
values of $k$ and $\nu$ approach the theoretical predictions of
Eq.~(\ref{exponents}), marked as dashed lines. We estimate the value of
$u_{\mathrm{c}}$ to be $u_{\mathrm{c}} = 1.92 \pm 0.03$ in $d=3$ and
$u_{\mathrm{c}}= 0.88 \pm 0.05$ in $d=2$ (marked by arrows).}
\label{nuk}
\end{figure}

\begin{figure}
\caption{Scaling plots for $\langle R^2(t) \rangle$~($\bigcirc$) 
and $\langle S(t) \rangle $~($\Box$) for $t \gg 1$ and 20 values
of $0 \le u \le 3$ in (a) $d=3$ and (b) $d=2$. For convenience, the data for 
$\langle S(t) \rangle $ have been shifted by $10^5$. In $d=3$ for
$\nu_{\mathrm{c}} = 0.32$,  $k_{\mathrm{c}} = 0.91$ and $u_{\mathrm{c}} = 1.92$
we find the best collapse for $\alpha = 5.0$, in $d=2$ for $\nu_{\mathrm{c}} =
0.40$,  $k_{\mathrm{c}} = 0.80$ and $u_{\mathrm{c}} = 0.88$ we find the best
collapse for $\alpha = 7$. The straight lines represent the exponents given in
Table~\ref{table}.}
\label{collaps3D}
\end{figure}


\begin{thebibliography}{MT1}
\bibitem{negative}
Note that, in the original definition by Sapozhnikov $p \sim \exp (-nu)$ with
$u<0$~\cite{Sapoz98}, for convenience we use $p \sim \exp (+nu)$ with $u>0$.
\bibitem{asymptotic}
These results could be found only because of the large number of
timesteps $t$ taken in our simulations, revealing the asymptotic behavior of
the exponents.
\end{thebibliography}
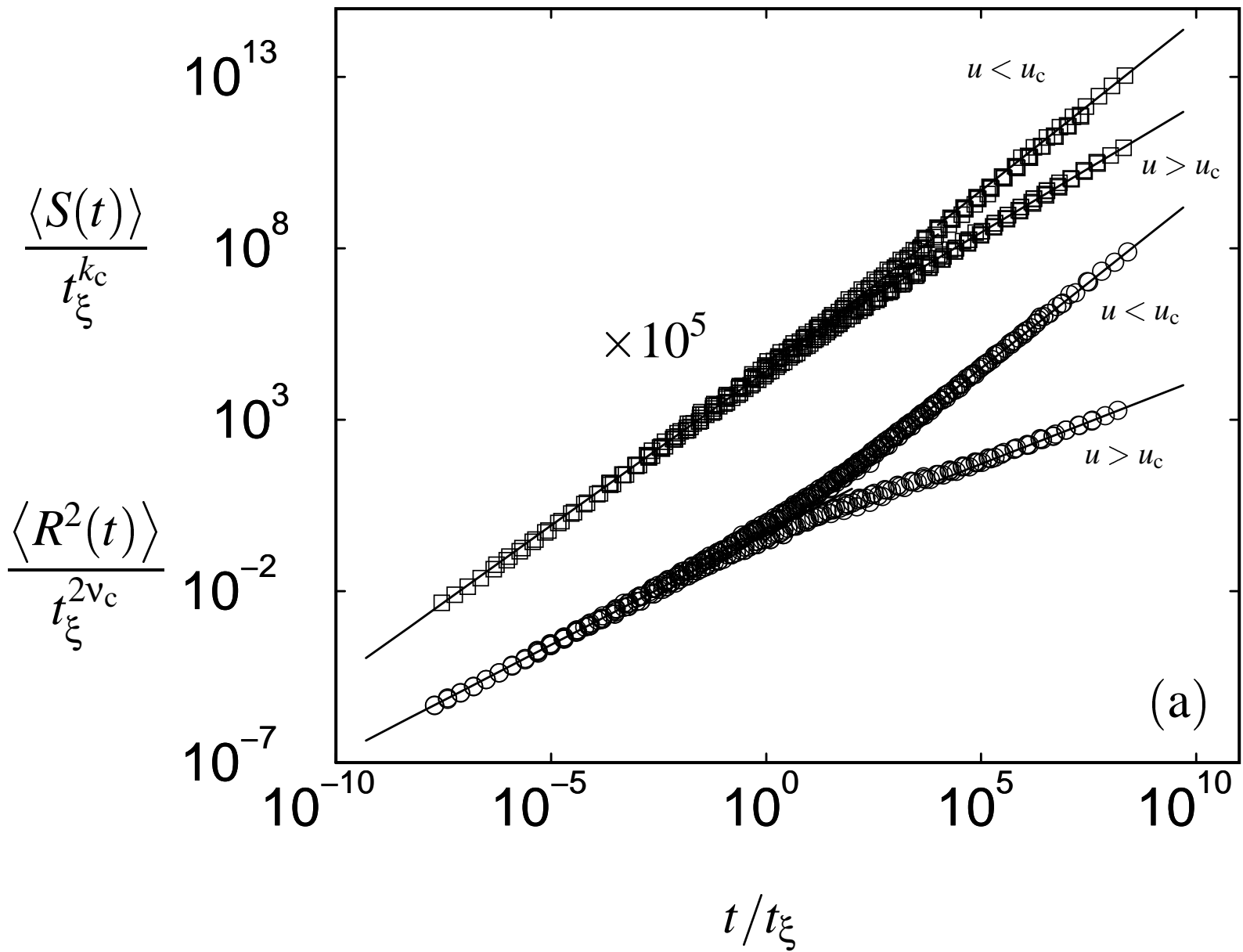
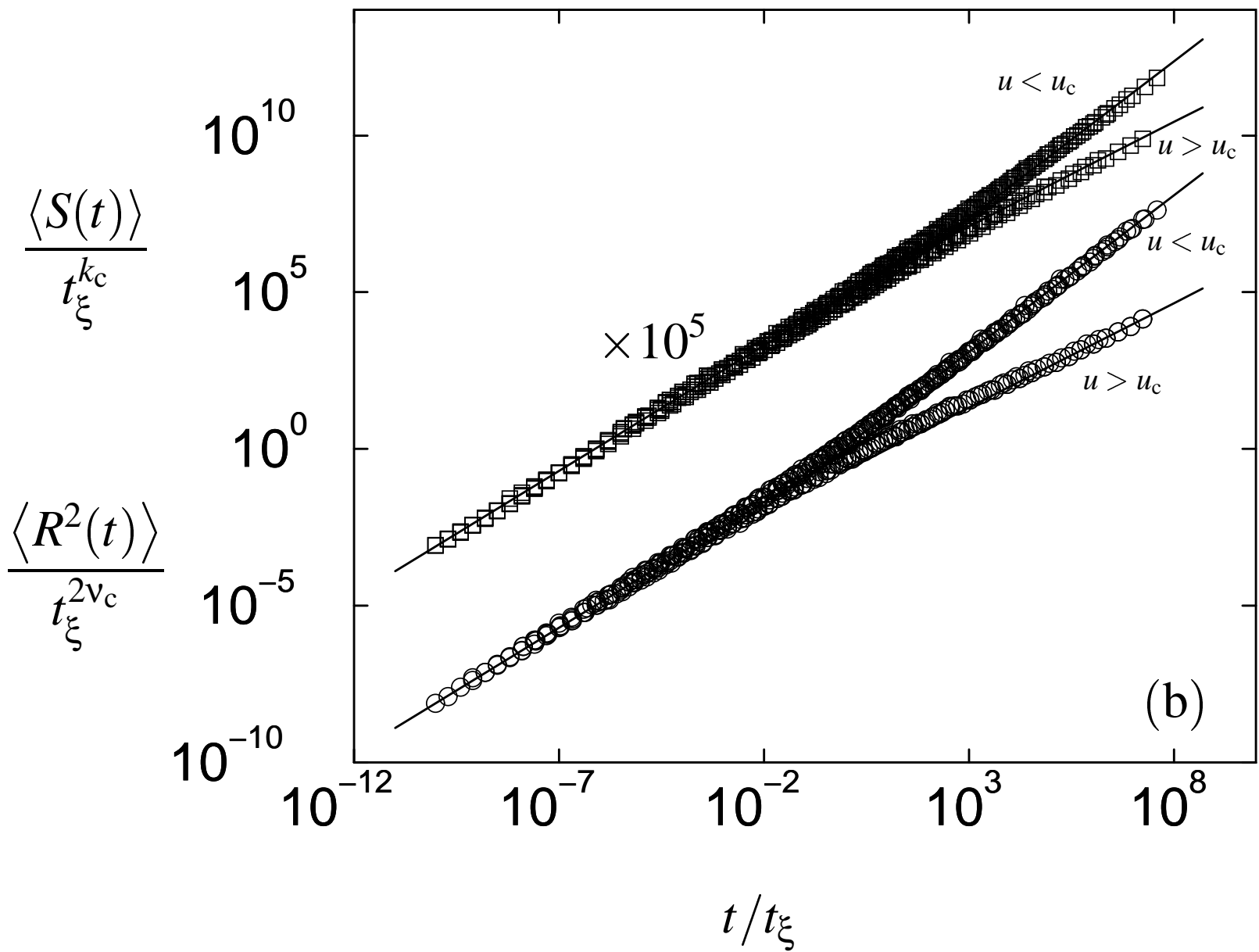
\end{document}